\documentclass[12pt,preprint]{aastex}

\shorttitle{$^{12}$CO(J = 2 -- 1) Imaging of Seyfert 1 Galaxy NGC 1097}

\shortauthors{Hsieh et al.}

\begin{document}

\title{Interferometric $^{12}$CO(J = 2 -- 1) image of
        the Nuclear Region of Seyfert 1 Galaxy NGC 1097}

\author{
        Pei-Ying Hsieh\altaffilmark{1,2},
        Satoki Matsushita\altaffilmark{2},
        Jeremy Lim\altaffilmark{2},
        Kotaro Kohno\altaffilmark{3},
        Satoko  Sawada-Satoh\altaffilmark{2,4}
\\pyhsieh@asiaa.sinica.edu.tw}

\affil{$^1$ Institute of Astrophysics, National Taiwan University,
		No.1 Sec. 4 Roosevelt Road, Taipei 10617, Taiwan, R.O.C.}
\affil{$^2$ Academia Sinica Institute of Astronomy and
		Astrophysics, P.O. Box 23-141, Taipei 10617, Taiwan, R.O.C.}
\affil{$^3$ Institute of Astronomy, The University of Tokyo,
		2-21-1 Osawa, Mitaka, Tokyo 181-0015, Japan}
\affil{$^4$ Department of Physics, Faculty of Science,
		Yamaguchi Univerity, 1677-1 Yoshida, Yamaguchi-City,
		Yamaguchi 752-8512 JAPAN}

\begin{abstract}
We have mapped the central region of the Seyfert 1 galaxy NGC 1097 in
$^{12}$CO(J = 2 -- 1) with the Submillieter Array (SMA).
The $^{12}$CO(J = 2 -- 1) map shows a central concentration and a
surrounding ring, which coincide respectively with the Seyfert
nucleus and a starburst ring.
The line intensity peaks at the nucleus, whereas in a previously
published $^{12}$CO(J = 1 -- 0) map the intensity peaks at the
starburst ring.
The molecular ring has an azimuthally averaged
$^{12}$CO(J = 2 -- 1)/(J = 1 -- 0) intensity ratio ($R_{21}$) of
about unity, which is similar to those in nearby active star forming
galaxies, suggesting that most of the molecular mass in the ring is
involved in fueling the starburst.
The molecular-gas-to-dynamical mass ratio in the starburst ring shows
a somewhat lower value than that found in nearby star forming
galaxies, suggesting that the high $R_{21}$ of unity may be caused by
additional effects, such as shocks induced by gas infall along the
bar.
The molecular gas can last for only about $1.2\times10^{8}$
years without further replenishment assuming a constant star formation
rate and a perfect conversion of gas to stars.
The velocity map shows that the central molecular gas is rotating
with the molecular ring in the same direction, while its velocity
gradient is much steeper than that of the ring.
This velocity gradient of the central gas is similar to what is
usually observed in some Seyfert 2 galaxies.
To view the active nucleus directly in the optical, the central
molecular gas structure can either be a low-inclined disk or torus
but not too low to be less massive than the mass of the host galaxy
itself, be a highly-inclined thin disk or clumpy and thick torus, or
be an inner part of the galactic disk.
The $R_{21}$ value of $\sim1.9$ of the central molecular gas component, which
is significantly higher than the value found at the molecular gas
ring, indicates that the activity of the Seyfert nucleus may have a
significant influence on the conditions of the molecular gas in the
central component.

\end{abstract}


\section{INTRODUCTION}
\label{sect-intro}

NGC 1097 \citep[SB(s)b;][]{dev91} is a nearby, inclined
\citep*[46$\degr$;][]{ondre89}, barred spiral galaxy
at a distance of D = 14.5 Mpc \citep[1\arcsec = 70 pc;][]{tully}.
It hosts a Seyfert 1 nucleus, which is surrounded by a circumnuclear
starburst ring with a radius of $10\arcsec$ (0.7 kpc).
NGC 1097 was originally identified as a LINER \citep{keel}, but over
the past two decades has shown Seyfert 1 activity as evidenced by the
presence of broad double-peaked Balmer emission lines
\citep*[FWHM $\thickapprox$ 7500 km s$^{-1}$;][]{SB93}.
The starburst ring is rich in molecular gas
\citep[$1.3\times10^{9}$ M$_{\odot}$;][]{ger88}, and exhibits a high
star formation rate of 5 M$_{\odot}$ yr$^{-1}$ as estimated from its
extinction-corrected H$\alpha$ luminosity \citep*{hum87}.
A network of dusty spiral features connects the starburst ring with
the nucleus \citep*{barth,pri05}.
Optical spectroscopic imaging observations show evidence for radial
streaming motions associated with the spiral structures leading to
the unresolved nucleus \citep{fat06}.

\citet{koh03} have previously mapped the central $\sim1\arcmin$ of
NGC 1097 in $^{12}$CO(J = 1 -- 0) and HCN(J = 1 -- 0) with the
Nobeyama Millimeter Array (NMA).
Their $^{12}$CO(J = 1 -- 0) map shows a ring-like structure consisting
of several bright knots located at the starburst ring, and this
molecular gas ring structure is similar to that revealed by the
single dish $^{12}$CO(J = 1 -- 0) observations \citep{ger88}.
The interferometric $^{12}$CO(J = 1 -- 0) map also revealed a
relatively weak central concentration coincident with the nucleus.
Their HCN(J = 1 -- 0) map shows similar knots in the starburst
ring, but in contrast to their $^{12}$CO(J = 1 -- 0) map
a relatively bright central concentration of the HCN emission
coincides with the nucleus.
This implies that the molecular gas associated with the nucleus is
relatively dense ($n_{\rm H_2} \ge 10^{4}$ cm$^{-2}$).
Similar dense central gas concentrations have been reported in
NGC 1068 \citep{jac93}, M51 \citep{koh96}, and NGC 6951
\citep{kri07}, all of which are type 2 Seyferts.
In all these cases, the dense central gas concentrations have been
attributed to a highly inclined circumnuclear molecular torus invoked
by AGN unification models \citep[e.g.,][]{ant93} to obscure the
central engine from direct view in the optical.
This is the first time, however, that such a dense central gas
concentration has been seen in a Seyfert 1 galaxy, warranting
observations at higher angular resolutions to determine whether it is
a good candidate for the hypothesized circumnuclear molecular torus.

In this paper, we study the central region of NGC 1097 in
$^{12}$CO(J = 2 -- 1) using the Submillimeter Array
\citep[SMA;][]{ho04}.
While the previous HCN(J = 1 -- 0) and $^{12}$CO(J = 1 -- 0)
interferometric observations are suitable for tracing density
variations in the molecular gas, they are not suitable for studying
variations in the gas temperature.
If the $^{12}$CO(J = 1 -- 0) line is optically thin, the intensity ratio
between the $^{12}$CO(J = 1 -- 0) line and the higher-J lines
provides constraints on the gas temperature (and density)
\citep*[e.g.,][]{mat04}.
In addition, we observe the $^{12}$CO(J = 2 -- 1) line at a factor of
2 -- 3 higher angular resolution than the previous observations in
$^{12}$CO(J = 1 -- 0) and HCN(J = 1 -- 0), allowing us to examine for
the first time the spatial-kinematic structure of the central gas
concentration.

\section{OBSERVATIONS AND DATA REDUCTION}
\label{sect-obs}

We observed the central region of NGC 1097 in the
$^{12}$CO(J = 2 -- 1) line (rest frequency of 230.538~GHz)
with the SMA, which has a primary beam of 52\arcsec~(3.6~kpc)
at this frequency.
The receivers were turned to observe the $^{12}$CO(J = 2 -- 1)
line at the upper side band.
The observations were performed on 2004 July 23 and October 1
with eight 6-m antennas in the compact configuration.
The 225 GHz zenith opacity of the two observations
are $\sim$0.15 and $\sim$0.3, and the ${T_{\rm sys, DSB}}$
are $\sim200$ K and $\sim350$ K respectively.
We placed the phase center at
$\alpha_{2000}$ = 02h46m18$\fs$96 and
$\delta_{2000}$ = -30${\degr}$16${\arcmin}$28${\farcs}$897,
which corresponds to the position of the AGN defined by the
intensity peak of the 6-cm continuum emission \citep{hum87}.
The SMA correlator has a total bandwidth of 2~GHz, and was configured
to provide frequency resolution of 0.8125~MHz ($\sim1$ km s$^{-1}$)
and 3.25~MHz (4.2 km s$^{-1}$) for the July 23 and October 1
observations respectively.
We observed Uranus for bandpass and absolute flux calibrations.
We used J0132--169 and J0423--013 for complex gain calibrations.
J0132--169 is weaker but closer to the source ($21\fdg5$ away), and
was used for phase calibration only; J0423--013 is stronger but
further from the source ($37\degr$ away), and used for amplitude
calibration.
The absolute flux accuracy and the positional accuracy have been
estimated as $\sim$15\% and $0\farcs1$, respectively, in these
observations.

The observation on July 23 was binned to the same velocity
resolutions as that on October 1 (3.25~MHz), and then combined with
each other.
The data were calibrated with the Owens Valley Radio Observatory
software package MIR, which was modified for the SMA.
The images were CLEANed using the NRAO software package AIPS, and
with natural weighting have an angular resolution of
$4\farcs1\times3\farcs1$ (290~pc $\times$ 220~pc) at a position angle
(P.A.) of $168\degr$.
The rms noise level of the individual channel maps is
75 mJy beam$^{-1}$.
We used the ``MOMNT'' task in AIPS to make the integrated intensity
and the intensity-weighted mean-velocity maps.
The line free channels were binned to create a continuum map with a
rms noise level of 4.4 mJy beam$^{-1}$;
no significant emission was detected in this map.

\section{RESULTS}
\label{sect-res}

We detected the $^{12}$CO(J = 2 -- 1) line spanning a total line
width of $\sim570$ km s$^{-1}$ at the $3\sigma$ detection threshold
in the channel maps.
The line width is comparable to that measured in the single-dish
observation \citep{pet03} of $\sim550$ km s$^{-1}$ using the JCMT,
which has a primary beam of $21\arcsec$, although the integrated
line intensity in our map is only $60\pm13\%$ of that measured in
this single-dish observation (see Sect.~\ref{sect-r21}) including
uncertainties of the flux in both observations.

\subsection{Spatial Distribution}
\label{sect-dist}

\begin{figure}
\epsscale{0.5}
\plotone{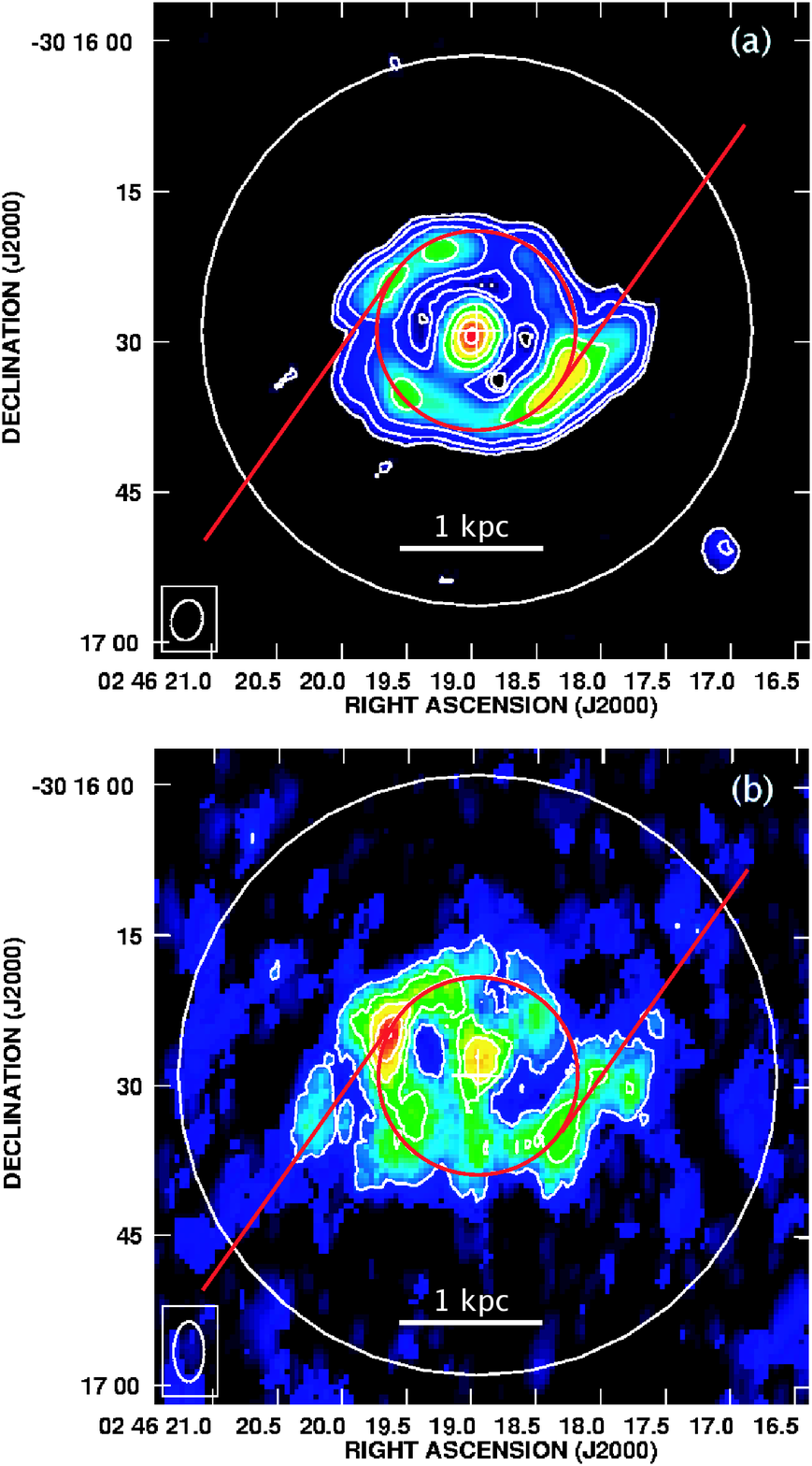}

\caption[]{(a) $^{12}$CO(J = 2 -- 1) integrated
intensity (moment 0) map of the central region of NGC 1097.
The contour levels are (2.5, 5, 10, 20, 30, 40, and 50) $\times$
3.7 Jy beam$^{-1}$ km s$^{-1}$, where 1$\sigma$ is 3.7
Jy beam$^{-1}$ km s$^{-1}$.
The synthesized beam is $3\farcs1\times4\farcs1$ with a P.A.\ of
$168\degr$, which is shown in the bottom-left corner of the map.
The large white circle is the primary beam size ($52\arcsec$) of the
SMA.
(b) $^{12}$CO(J = 1 -- 0) integrated intensity map.
The contour levels are (2.5, 5, 7, 9, and 11) $\times$
3.9 Jy beam$^{-1}$ km s$^{-1}$, where 1$\sigma$ is
3.9 Jy beam$^{-1}$ km s$^{-1}$.
The synthesized beam is $6\farcs0\times3\farcs0$ with a P.A.\ of
$0\degr$, which is shown in the bottom-left corner.
The large white circle is the primary beam size ($63\arcsec$) of the
NMA.
In both images, the central crosses mark the peak of the 6~cm
continuum \citep{hum87}, the small red circles at a radius of
$10\arcsec$ mark the radius of the molecular ring, and the two red
lines mark the positions of the dust lanes.
\label{fig-intmap}}
\end{figure}

We show the $^{12}$CO(J = 2 -- 1) integrated intensity map in
Figure~\ref{fig-intmap}(a).
A relatively strong molecular concentration (hereafter, central
component) is detected at the very center in a region of a diameter
of 350~pc ($5\arcsec$), surrounded by a weaker molecular ring of
a diameter of 1.4~kpc ($20\arcsec$).
The centroid of the central molecular gas component is shifted by
$\sim0\farcs7$ from that of the radio continuum core at 6~cm, which
presumably marks the location of the AGN \citep{hum87}.
We shall return to this positional difference in
Section~\ref{sect-kin}.

The molecular ring coincides with the starburst ring, consistent with
the previous $^{12}$CO(J = 1 -- 0) maps \citep{ger88,koh03}.
Unlike the tightly-wound spiral structure seen in optical maps
\citep[e.g.,][]{rick,barth}, the molecular ring traces a complete
circle as seen in the $J-K_{S}$ color map of \citet{pri05}.
The molecular ring is composed of several knots that closely
resembles the 1.5 GHz continuum map of \citet{hum87}.
In addition, we detected weak molecular emission extending from the
NE and SW of the molecular ring, coinciding with dust lanes along the
large stellar bar (indicated by the two straight lines in
Fig.~\ref{fig-intmap}a).

In Figure~\ref{fig-intmap}(b), we show the $^{12}$CO(J = 1 -- 0) map
made by \citet{koh03} with the NMA.
As can be seen, the molecular ring is more clearly discernible and
also more clearly separated in our $^{12}$CO(J = 2 -- 1) map than in
the $^{12}$CO(J = 1 -- 0) map.
The intensity peak in the $^{12}$CO(J = 2 -- 1) map is
located at the nucleus, whereas that in the $^{12}$CO(J = 1 -- 0)
map is located at the NE part of the ring.
The SW side of the ring is brighter than the NE side in the
$^{12}$CO(J = 2 -- 1) map, but the reverse is true in the
$^{12}$CO(J = 1 -- 0) map.
Both of the molecular gas peaks at the NE and SW sides of the ring
are located at the regions where the dust lanes connect with the
ring, and these molecular gas peaks are often seen in barred spiral
galaxies, which are known as the twin-peak morphology
\citep[e.g.,][]{ken92}.

\subsection{Kinematics}
\label{sect-kin}

\begin{figure}
\epsscale{0.4}
\plotone{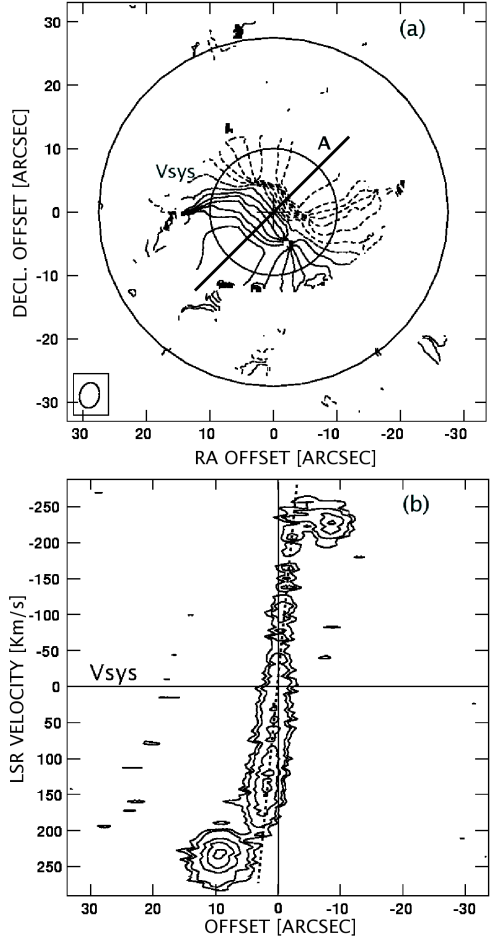}

\caption[]{(a) Intensity-weighted mean velocity
(moment 1) map of $^{12}$CO(J = 2 -- 1).  Large and small circles
are marked as the primary beam size of the SMA ($52\arcsec$) and the
size of starburst ring (radius of $10\arcsec$), respectively.
The systemic velocity \citep[1254 km s$^{-1}$;][]{koh03} is marked.
The contour interval is 15 km s$^{-1}$.
Dotted and solid curves are the blueshifted and redshifted
velocities, respectively.
The solid line labeled as ``A'' represents the direction of the major
axis (P.A.\ is 135$\degr$) of NGC 1097 \citep{hum87}.
The nuclear gas is rotating with the molecular ring in the same
sense, while it is rotating faster than the ring.
(b) Position-velocity diagram of the $^{12}$CO(J = 2 -- 1) emission
cut along the major axis (represented as ``A'' in the above figure).
Contour levels are 3, 5, 10, 15, and $20\sigma$, where
$1\sigma$ = 75 mJy beam$^{-1}$.
The zero velocity corresponds to the systemic velocity of
1254 km s$^{-1}$ marked as horizontal solid lines.
The offset of $0\arcsec$ corresponds to the galactic center, and the
LSR velocity of 0 km s$^{-1}$ corresponds to the systemic velocity.
The fitted velocity gradient (1.31 km s$^{-1}$) of the
nuclear component is marked as a dash line.
Note that the inclination of the galactic disk is not corrected.}
\label{fig-pvmom1}
\end{figure}

In Figure~\ref{fig-pvmom1}(a), we show the intensity-weighted
$^{12}$CO(J = 2 -- 1) mean-velocity map.
Both the central component and molecular ring show an overall
velocity gradient in the NE to SE direction along a P.A.\ of
$135\degr$, that is similar to the major kinematic axis of the
large-scale galactic disk \citep{ondre89}.
The emission is blueshifted on the northwestern side of center, and
redshifted on the southeastern side of center with respect to the
systemic velocity of 1254 km s$^{-1}$ for NGC 1097 \citep{koh03}.
The gas motion in the molecular ring appears to be dominated by
circular motion (i.e., isovelocity contours perpendicular to major
axis), whereas in the $^{12}$CO(J = 1 -- 0) velocity map of
\citet{koh03} the isovelocity contours
have a symmetric S-shape with the end of the S-shape nearly parallel
to the dust lanes along the large stellar bar.
This non-circular motion is not as prominent in our
$^{12}$CO(J = 2 -- 1) map perhaps because the aforementioned dust
lanes are not as strongly detected in $^{12}$CO(J = 2 -- 1), along
with the fact that they are closer to the edge of our primary beam
than in the $^{12}$CO(J = 1 -- 0) observations of \citet{koh03}.

A position-velocity diagram (hereafter p-v diagram) of the
$^{12}$CO(J = 2 -- 1) emission along its major kinematical axis
(P.A.\ = $135\degr$) is shown in Figure~\ref{fig-pvmom1}(b).
Positive and negative velocities correspond to redshifted and
blueshifted velocities, respectively, relative to the systemic
velocity.
The rapidly rising part of the rotation curve corresponds to the
central component, and the flat part to the molecular ring.
The rotation curve is symmetric on both sides of the center, rising
steeply to $\pm235$ km s$^{-1}$ at $\pm2\farcs5$ ($\pm175$~pc) and
flattening outside $\pm2\farcs5$.
This indicates that the size of the central component is about 350~pc
in diameter.
As can be seen in Figure~\ref{fig-pvmom1}(b), the emission from the
central component is stronger on the redshifted southeastern part,
causing the centroid of the emission to be shifted by $0\farcs$7
towards the southeast as mention in Section~\ref{sect-dist} (and
shown in Fig.~\ref{fig-intmap}a).
The velocity gradient of the central component is
$1.31\pm0.14$ km s$^{-1}$ pc$^{-1}$ (Fig.~\ref{fig-intmap}b), which
is similar with that derived in the optical from ionized gas
\citep*[$\sim$ 1.1 km s$^{-1}$ pc$^{-1}$;][]{sto96}.

\subsection{$^{12}$CO(J = 2 -- 1)/(J = 1 -- 0) Line Ratios}
\label{sect-r21}

Our $^{12}$CO(J = 2 -- 1) map detected about $60\%$ of the flux
measured with the JCMT \citep{pet03}.
The shortest projected baseline in our SMA observations is
4.8~k$\lambda$ ($\sim6$~m), and so the largest structure we can
detect is $\sim$52$\arcsec$, comparable in size to the SMA primary
beam.
We therefore assume that the molecular gas resolved out in our map
is uniformly distributed in space and in velocity over the entire
line width, and have corrected for $\sim40\%$ of the missing flux.
We applied a primary beam correction to both the
$^{12}$CO(J = 2 -- 1) and $^{12}$CO(J = 1 -- 0) maps, and convolved
the derived moment maps to the same angular resolution of
$6\farcs5\times3\farcs5$ with P.A.\ of $0\degr$.
These moment maps were then used to derive the
$^{12}$CO(J = 2 -- 1)/$^{12}$CO(J = 1 -- 0) line ratios.
Note that we also calculated the line ratios with $uv$-matched data,
and the ratios are consitent with the missing flux corrected line
ratios within errors.
To derive the molecular gas column densities and masses, we need to
use the missing flux corrected data (see Sect.~\ref{sect-phys}).
To make consistency within this paper, we use the missing flux
corrected data for the following calculations.

\begin{figure}
\epsscale{1}
\plotone{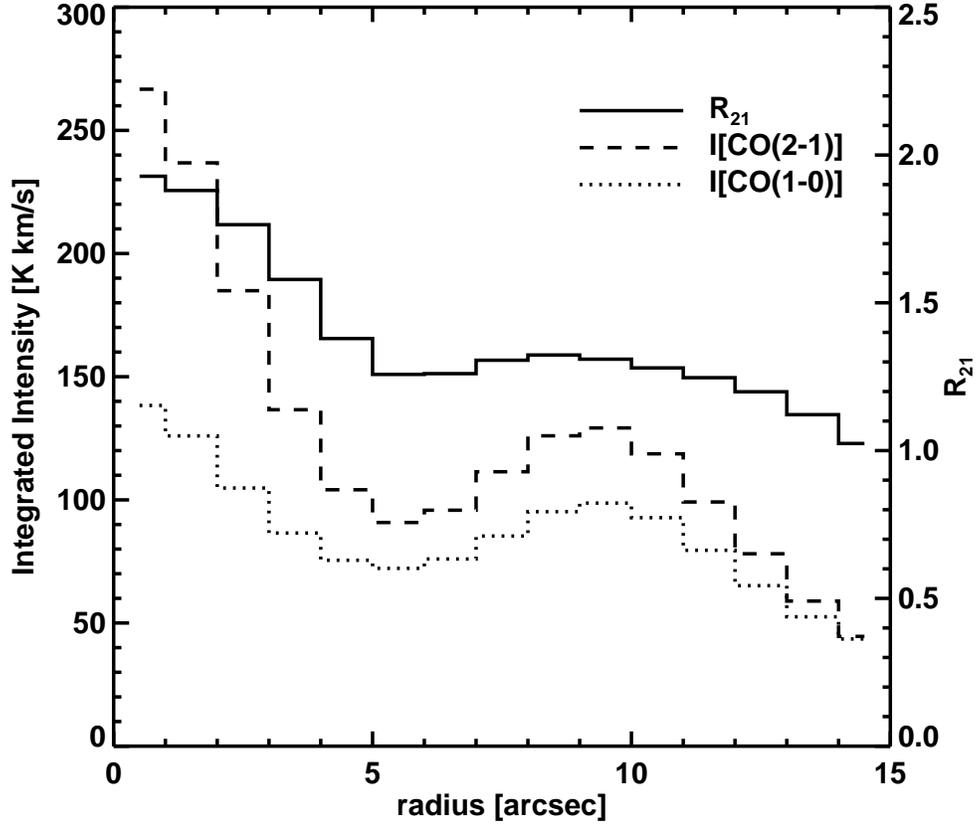}

\caption[]{Azimuthally averaged radial intensity distributions of
the $^{12}$CO(J = 2 -- 1) (dashed line) and $^{12}$CO(J = 1 -- 0)
(dotted line) lines in brightness temperature scale.
The temperature scale is shown in the left-hand side of the
vertical axis. Azimuthally averaged radial distribution of
$R_{21}$ is marked as a solid line. The ratio scale is shown
in the right-hand side of the vertical axis.}
\label{fig-ratioaz}
\end{figure}

In Figure~\ref{fig-ratioaz}, we show the azimuthally averaged
radial intensity distributions of the $^{12}$CO(J = 2 -- 1) (dashed
line) and $^{12}$CO(J = 1 -- 0) (dotted line) lines, and the
azimuthally averaged $^{12}$CO(J = 2 -- 1)/$^{12}$CO(J = 1 -- 0) line
intensity ratio ($R_{21}$)(solid line).
The radial intensity profile in both $^{12}$CO(J = 2 -- 1) and
$^{12}$CO(J = 1 -- 0) peak at the center, corresponding to the
central component, and exhibit a secondary peak at a radius of
$10\arcsec$, corresponding to the molecular ring.
The line ratio $R_{21}$ exhibits a similar behavior, peaking at the
central component and exhibiting a secondary peak at the molecular
ring.
At the nucleus, $R_{21}$ is derived as $1.9\pm0.2$ at a beam size of
$6\farcs5\times3\farcs5$, which corresponds to the approximate
angular extent of the central component along its major kinematic
axis.
In the molecular ring, the azimuthally averaged value between radii
of 8$\arcsec$ and 12$\arcsec$ is $R_{21} = 1.3\pm0.2$.

The averaged $R_{21}$ at the molecular ring is similar to or somewhat
larger than the global $R_{21}$ in spiral galaxies of $\sim0.7-0.9$
\citep{bra93,lav99,haf03}.
The $R_{21}$ at giant molecular clouds (GMCs) in the nearby star
forming region Orion \citep{sak94} or GMC-scale $R_{21}$ in the
nearby starburst galaxy M82 \citep{wei01} also shows similar $R_{21}$
of about unity.
On the other hand, the $R_{21}$ at the central component is about
twice as high as those in the abovementioned sources, and consistent
with the ratio reported in the Seyfert 2 galaxy NGC 1068
\citep{baker98,sch00}.
The $R_{21}$ at the central component is also similar to that found at
active star forming regions or at the interfaces of molecular clouds
and ionized gas in the nearby barred spiral galaxy IC 342
\citep*{tur93,mei00}.
Galactic objects, such as molecular outflows from young stellar
objects \citep[e.g.,][]{ric85,cha96} or molecular gas around
supernova remnants \citep*{van93,set98} also show $R_{21}\gtrsim2$.
Note however that the high value of $R_{21}$ are not always seen in
active galaxies; for instance, interferometric observations of
NGC 3227 \citep*[Seyfert 1;][]{sch2000},
NGC 3718 \citep[LINER/Seyfert 1;][]{kri05}, and
NGC 6574 \citep[Seyfert 2;][]{lindt08} show $R_{21}$ of around unity.

\subsection{Physical Properties of the Molecular Gas}
\label{sect-phys}

We derived the column density of the molecular gas in the central
component and molecular ring assuming the standard Galactic
conversion factor of
$3.0\times10^{20}$ cm$^{-2}$ (K km s$^{-1}$)$^{-1}$ between the
$^{12}$CO(J = 1 -- 0) line intensity and column density of molecular
hydrogen gas \citep{sco87,sol87}.
The $^{12}$CO(J = 2 -- 1) integrated intensity at the center (beam
size of $3\farcs1\times4\farcs1$) is 376.7$\pm$7.2 K km s$^{-1}$.
Assuming the same line ratios derived from a beam size of
$6\farcs5\times3\farcs5$, namely, $R_{21} = 1.9\pm0.2$ for the
central component, we derive a column density for this component of
$(5.9\pm0.6)\times10^{22}$ cm$^{-2}$, and the molecular hydrogen gas
mass of $(6.5\pm0.7)\times10^{7}~{\rm M_{\odot}}$.
In the molecular ring, the azimuthally averaged $^{12}$CO(J = 2 -- 1)
integrated intensity between the radii of $8\arcsec$ and $12\arcsec$
is $135.7\pm7.2$ K km s$^{-1}$.
With $R_{21} = 1.3\pm0.2$, the azimuthally averaged column density at
the molecular ring is therefore $(3.0\pm0.3)\times10^{22}$ cm$^{-2}$,
and the molecular hydrogen gas mass is
$(5.8\pm0.6)\times10^{8}~{\rm M_{\odot}}$.
Here we compare our estimated molecular gas mass of the ring with
that estimated using the single-dish data.
We re-calculated the gas mass inside 20$\arcsec$ in radius derived by
\citet{ger88} using the conversion factor mentioned above, and it is
calculated as $1.1\times10^{9}~{\rm M_{\odot}}$.
We then subtract the gas mass of the center from this gas mass.
The gas mass of the center is calculated as
$2.4\times10^{8}~{\rm M_{\odot}}$ using the central position data
with a $T_{\rm b}$ of $\sim$32 K km s$^{-1}$
(Fig.~3 of \citet{ger88}).
The gas mass of the ring is therefore calculated as
$\sim$$8.6\times10^{8}~{\rm M_{\odot}}$.
This value is somewhat larger than ours, and this may be due to their
larger radius of 20$\arcsec$, and therefore their results may be
detecting emission outside the ring.

To determine the temperature and density of the molecular gas, we use
the LVG method \citep{gold74}.
We assume a one zone model, which assumes that both the
$^{12}$CO(J = 2 -- 1) and $^{12}$CO(J = 1 -- 0) emissions originate
from the same region.
The collision rates for CO are taken from \citet{flo85} for the
temperature regime T = $10-250$~K, and \citet{lvg-h-t} for
T = $500-1000$~K.
We first assume a standard $Z(^{12}$CO)/(d$v$/d$r$) of
$5.0\times10^{-5}$ (km s$^{-1}$ pc$^{-1}$)$^{-1}$, where
$Z(^{12}$CO) = [$^{12}$CO]/[H$_{2}$] is the abundance ratio and
d$v$/d$r$ is the velocity gradient of the molecular gas.

\begin{figure}
\epsscale{0.65}
\plotone{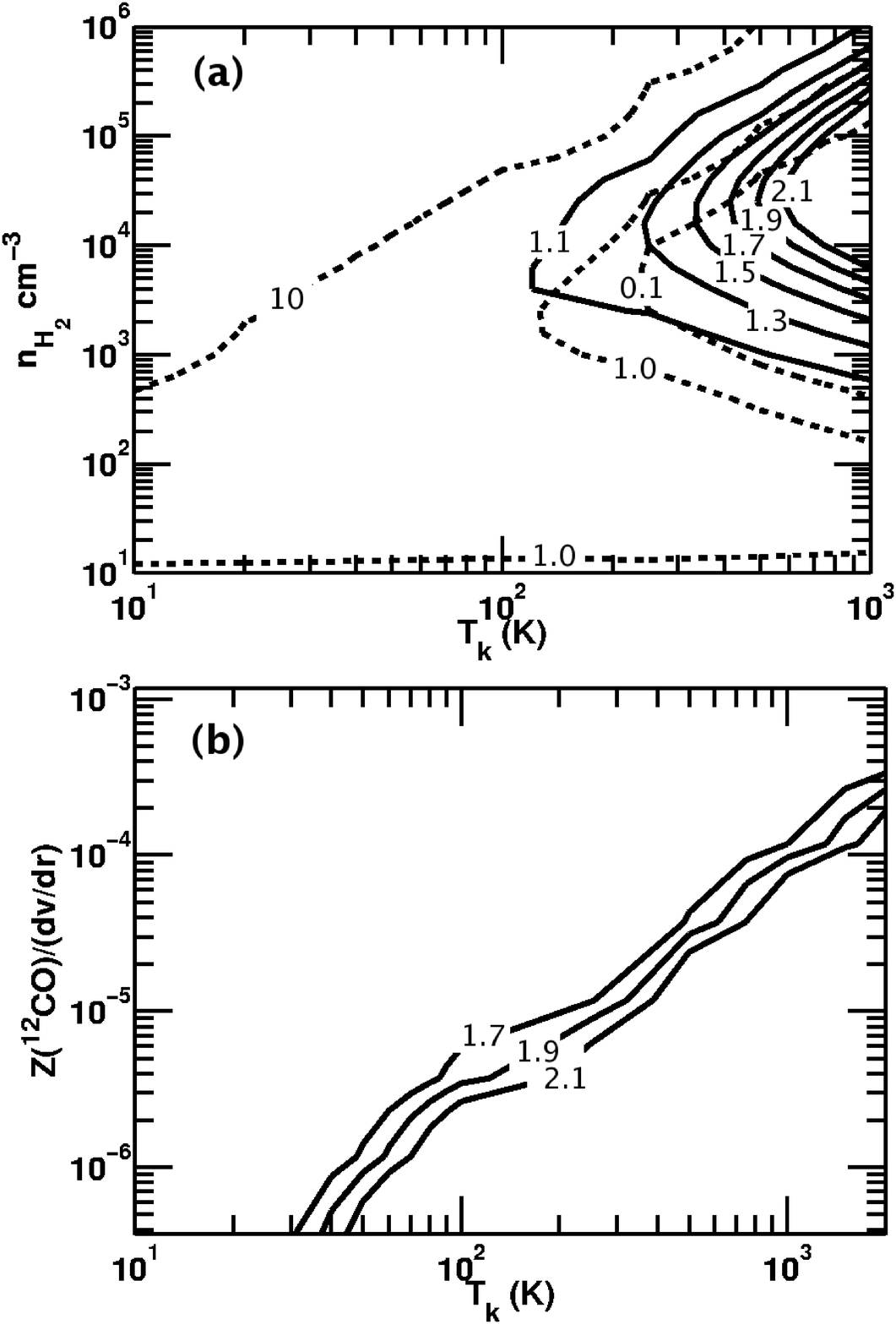}

\caption[]{(a) $R_{21}$ is plotted as a function of kinetic
temperature $T_{\rm k}$ and H$_{2}$ number density $n_{\rm H_2}$ at a
standard $Z(^{12}$CO)/(d$v$/d$r$) of
$5.0\times10^{-5}$ (km s$^{-1}$ pc$^{-1}$)$^{-1}$.
Solid lines are the $R_{21}$ of the nuclear component (1.9$\pm$0.2)
and the molecular ring (1.3$\pm$0.2).
Opacity for $^{12}$CO(J = 1 -- 0) is marked as dashed lines.
(b) $R_{21}$ of the nuclear component is plotted as a function of
$T_{\rm k}$ and $Z(^{12}$CO)/(d$v$/d$r$).
Solid lines are the observed $R_{21}$ of the nuclear component with
$n_{\rm H_2}$ fixed at a value of $1.0\times10^{4}$ cm$^{-3}$.
\label{fig-tknh2}}
\end{figure}

We then compute from the LVG model the line ratio $R_{21}$ and
$^{12}$CO(J = 1 -- 0) opacity as a function of molecular hydrogen
number density $n_{\rm H_2}$ and kinetic temperature $T_{\rm k}$ as
shown in Figure~\ref{fig-tknh2}(a).
The result for $R_{21}=1.9\pm0.2$ as inferred for the central
component indicates that $T_{\rm k}\ge400$~K, and
$n_{\rm H_2}\sim3\times10^{4\pm1}$~cm$^{-3}$, and
$^{12}$CO(J = 1 -- 0) opacity below unity.
The density estimation is consistent with the detection of
HCN(J = 1 -- 0) line from the central component \citep{koh03}, which
indicates the molecular gas density as high as
$n_{\rm H_2}\approx10^{4}$~cm$^{-3}$.
Note that the estimated kinetic temperature is highly dependent on
the assumed $Z(^{12}$CO)/(d$v$/d$r$), namely highly dependent on the
[$^{12}$CO]/[H$_{2}$] relative abundance, on the velocity gradient,
or on both.
Fixing $n_{\rm H_2}=1\times10^{4}$~cm$^{-3}$, we plot $R_{\rm 21}$ as
a function of $T_{\rm k}$ and $Z(^{12}$CO)/(d$v$/d$r$) in
Figure~\ref{fig-tknh2}(b).
As can be seen, $R_{21}$ increases roughly linearly with
$Z(^{12}$CO)/(d$v$/d$r$), and around the standard
$Z(^{12}$CO)/(d$v$/d$r$) of $\approx$10$^{-5}$
(km s$^{-1}$ pc$^{-1}$)$^{-1}$, we find that a kinetic temperature of
at least 100~K is required to reach $R_{\rm 21}$ of $1.9\pm0.2$.
If on the other hand $Z(^{12}$CO)/(d$v$/d$r$) is an order of
magnitude lower than the standard value (i.e., order of 10$^{-6}$),
the kinetic temperature would be in the range of
$T_{\rm k}$ $\approx$ $30-250$~K, which is comparable with the
temperature range normally found in molecular clouds.

For the molecular ring, which has an average $R_{21}=1.3\pm0.2$,
$T_{\rm k}\gtrsim100$~K and
$n_{\rm H_2}\sim1\times10^{4\pm1}$~cm$^{-3}$
assuming the standard $Z(^{12}$CO)/(d$v$/d$r$).
Again, these values are sensitive to the assumed
$Z(^{12}$CO)/(d$v$/d$r$), and if this value decreases, $T_{\rm k}$
and $n_{\rm H_2}$ also decrease.

\section{DISCUSSIONS}
\label{sect-dis}

\subsection{Molecular Gas in the Starburst Ring}
\label{sect-dis-ring}

The average $R_{21}$ of about unity in the molecular ring suggests
that the overall properties of the molecular gas in the starburst
ring is optically thick and thermalized.
As mentioned in Section~\ref{sect-r21}, this value is similar to that
of the global $R_{21}$ of other spiral galaxies or that in the
active star forming GMCs.
Similarity of the molecular gas ratios between the molecular gas
in the ring and the molecular clouds at active star forming regions
indicates that the molecular gas in the ring is closely related to
the starburst activities occurring in the ring.
This is also supported by the high
HCN(J = 1 -- 0)/$^{12}$CO(J = 1 -- 0) of 0.16 \citep{koh03}.
Since stars form from molecular gas, we believe that this molecular
gas is actually fueling the starburst activities in the ring.
The gas mass content in the starburst ring is, on the other hand, not
as high as star forming galaxies, about $8\%$ of the dynamical mass;
the dynamical mass within the radius of $8\arcsec$ -- $12\arcsec$
(560 -- 840~pc; roughly corresponds to the width of the molecular
ring) can be estimated as $(6.9\pm0.2)\times10^{9}~{\rm M}_\odot$,
using the rotational velocity of 235 km s$^{-1}$
(Sect.~\ref{sect-kin}) and assuming the inclination of the molecular
ring is the same as that of the host galaxy of $46\degr$
\citep{ondre89}.
The molecular gas mass in the ring is $5.8\times10^{8}~{\rm M}_\odot$
(Sect.~\ref{sect-phys}), so that the molecular gas mass to dynamical
mass ratio is calculated as $(8\pm1)\%$.
This value is a bit lower than the threshold to have star formation
activities at nuclei of $\sim10\%$ suggested by \citet{sak99}.
One reason for the discrepancy between high $R_{21}$ and
HCN(J = 1 -- 0)/$^{12}$CO(J = 1 -- 0) ratios and low molecular gas
mass to dynamical mass ratio may be that \citet{sak99} are taking
mass ratios within 500~pc in radius, but we are taking the mass ratio
in a certain range of radius.
Another possibilities is that the gas in the staburst ring is highly
affected by shock induced by the gas infall along the bar, so that
the high intensity ratios may not tracing the star formation
activities, but tracing shocked gas.

The upper limit of the mass of newly formed (1.5~Gyr old) stars in
the star-forming ring is $\sim10^{9}~{\rm M}_\odot$ \citep{qui95},
which is roughly comparable to or somewhat larger than the mass of
molecular gas ($5.8\times10^{8}~{\rm M}_\odot$).
This implies that a large fraction of the molecular gas has already
turned into stars within a timescale of 1.5~Gyr.
We can estimate the molecular gas consumption timescale with
considering the star formation rate (SFR) and the molecular gas mass
of the ring.
The SFR in the molecular ring is about 5~M$_\odot$ yr$^{-1}$
\citep{hum87}.
Without further replenishment, the molecular gas in the ring can last
for only about $(1.2\pm0.1)\times10^{8}$ years assuming a constant SFR
and a perfect conversion of gas to stars.
This value is roughly an order of magnitude lower than the value
derived by \citet{ger88}, which is due to their somewhat larger
molecular gas mass, inclusion of atomic hydrogen gas mass assuming it
has the same amount as the molecular gas mass, and using lower SFR of
$\sim2$~M$_\odot$ yr$^{-1}$.
In addition to the gas consumption by star formation, the amount of
molecular gas in the ring also decreases by the infall toward the
nucleus.
As mentioned in Section~\ref{sect-intro}, the gas infalling motion has
been observed from the ring to the nucleus.
The value for the mass infall rate is not known, but this effect
obviously decreases the lifetime of the molecular ring.
On the other hand, gas replenishment by the infall along the large
scale bar from the outer part of the galaxy to the molecular ring
increases the lifetime.
The value for the mass infall rate is again not known, so the
lifetime of the molecular ring is decided by the balance between
these effects.

\subsection{Nature of the Nuclear Component}
\label{sect-dis-nature}

Our $^{12}$CO(J = 2 -- 1) results show that the molecular gas
located at the center is spatially coincident with the AGN, which is
consistent with the previous interferometric maps of the
$^{12}$CO(J = 1 -- 0) and HCN(J = 1 -- 0) lines \citep{koh03}.
In addition, the central component is kinematically symmetric at the
center with steep velocity gradient, suggesting that the molecular
gas is rotating around the AGN.
Here we discuss the relation between the central component and the
AGN activities based on the line ratios of the central component and
its physical conditions, which were derived using the LVG analysis
results presented in the previous section.

The $^{12}$CO(J = 2 -- 1) map peaks at the nucleus, different from
the $^{12}$CO(J = 1 -- 0) map, and the $R_{21}$ shows $1.9\pm0.2$.
This value is significantly higher value than the global $R_{21}$ of
spiral galaxies or $R_{21}$ in star forming GMCs in Orion or M82, but
similar to the $R_{21}$ of molecular gas at jets, at supernova
remnants, or at the surface between molecular gas and ionized gas
(Sect.~\ref{sect-r21}).
These results suggest that the high $R_{21}$ in molecular gas can be
related to irradiation of UV photons from star forming regions, shock
caused by interaction between molecular gas and outflowing or
expanding materials, but not to the global galactic characteristics
or activities.
Hence similar activities can be the cause of high $R_{21}$ in the
central region of NGC 1097.
However, other activities that cannot be seen in star forming
galaxies or in our Galaxy can also be the cause of high $R_{21}$,
such as strong X-ray radiation from the Seyfert 1 nucleus
\citep*[e.g.,][]{lepp96,koh01,usero04,koh05,mei07}.

From LVG analysis, high temperature ($\gtrsim400$~K) and high density
($\sim3\times10^{4\pm1}$~cm$^{-3}$) conditions are derived for the
central component, assuming a standard $Z(^{12}$CO)/(d$v$/d$r$).
As mentioned in Section~\ref{sect-phys}, the density is consistent
with the centrally peaked HCN(1 -- 0) map \citep{koh03}.
Furthermore, these conditions are supported by results from infrared
observations:
Strong molecular hydrogen line, H$_{2}$ 1 -- 0 S(1), is detected
toward the nucleus without any detection of Br$\gamma$ line, and the
$JHK$ images show red colors toward the nucleus, suggesting that the
presence of hot dust coexisting with the dense molecular gas
\citep{kot00}.
Such high density and high temperature conditions for molecular gas
around Seyfert nuclei are also derived in other galaxies, such as
the Seyfert 2 nucleus of M51 \citep{mat98,mat04} or NGC 1068
\citep{rot91,tac94}, but not
for the molecular gas in non-active galaxies such as IC 342 or our
Galaxy \citep{mat98}.
These physical values are, however, sensitive to
$Z(^{12}$CO)/(d$v$/d$r$), namely to the $^{12}$CO abundances, to
the velocity gradient, or to both.
As shown in Section~\ref{sect-phys},
lower $Z(^{12}$CO)/(d$v$/d$r$) makes the derived temperature lower.
This is because the opacity is linearly related to
$Z(^{12}$CO)/(d$v$/d$r$), and it will decrease if
$Z(^{12}$CO)/(d$v$/d$r$) decreases.
Under low $Z(^{12}$CO)/(d$v$/d$r$) conditions, if the temperature
rises,
the $^{12}$CO(J = 1 -- 0) line can easily be optically thin, and easier
to have high $R_{21}$ at lower temperature than the normal
$Z(^{12}$CO)/(d$v$/d$r$) condition.
In either case, the molecular gas around Seyfert nuclei seems to have
different properties from other normal environment, which seems to be
largely related to the Seyfert activities.
Combined with the spatial and kinematical information, we suggest
that the central component is kinematically and physically related to
the Seyfert nucleus.

\subsection{Is the Circumnuclear Gas the Hypothetical Circumnuclear
    Torus?}
\label{sect-dis-torus}

Given that the central component is closely related to the AGN, is
this then the circumnuclear molecular torus predicted by AGN unified
models?
Since NGC 1097 hosts a Seyfert 1 AGN, we expect that the rotating
circumnuclear gas is in a nearly face-on configuration to provide an
essentially unobscured view to the broad-line region (BLR),
if we apply the general unified model for AGNs
\citep[e.g.,][]{ant93}.
Kinematics of the central component we observed are, however, similar
to what is usually observed in other Seyfert 2 galaxies, such as
NGC 1068 \citep{jac93,sch00} or M51 \citep{koh96,sco98}; the central
component shows steep velocity gradient that can be explained by
edge-on disk or torus rotating around the Seyfert nucleus.
In addition, we derived a molecular hydrogen column density
($N_{\rm H_2}$) toward the nucleus of
$(5.9\pm0.6)\times10^{22}$ cm$^{-2}$ (Sect.~\ref{sect-phys}), or
$1.2\times10^{23}$ cm$^{-2}$ in atomic hydrogen column density
($N_{\rm H}$), which is about two orders of magnitudes larger than
$N_{\rm H}$ derived from X-ray spectra of
$1.3\times10^{21}$ cm$^{-2}$ \citep{iyo96,tera02}.

From these ``inconsistent'' results between our molecular gas
observations and other observations at different
wavelengths/frequencies, the structure of the central component can
have two basic configurations; one is a nearly face-on disk-, ring-,
or torus-like structure with a very fast rotation velocity, and
another is a nearly edge-on disk-, ring-, or torus-like structure
with a thin or clumpy structure.
Due to our large synthesized beam size, we could not distinguish
these two possibilities observationally, and also could not evaluate
the thickness of the structure.
Here we discuss the advantages and disadvantages of both
possibilities.

The former molecular gas configuration has a nearly face-on
structure, so the direct view to the BLR is secured.
But since the observed rotational velocity width has
470 km s$^{-1}$, the inclination corrected rotational velocity width
for this configuration has to be $470\sin(i)$, where $i$ is the
inclination ($0\degr$ corresponds to a face-on configuration).
The nearly face-on configuration therefore leads to a high rotational
velocity (e.g., 940 km s$^{-1}$ even for the inclination angle of
$30\degr$).
Note that this rotation is rigid rotation, and it is rare to see
rigid rotation velocity of $>500$ km s$^{-1}$ in other galaxies
\citep*[e.g.,][]{rubin1,rubin2,sofue}.
In addition, assuming the inclination angles as $30\degr$, $10\degr$,
and $5\degr$, the dynamical mass within the central gas, namely
within the radius of 2$\farcs$5 (175~pc), can be estimated to be 
$8.8\times10^{9}$$~{\rm M_{\odot}}$,
$7.3\times10^{10}$$~{\rm M_{\odot}}$, and
$2.9\times10^{11}~{\rm M_{\odot}}$, respectively.
The total dynamical mass of NGC 1097 within a radius of 7$\farcm$5
(31.5~kpc) estimated from the large-scale atomic hydrogen
observations is $(5.0\pm0.8)\times10^{11}~{\rm M_{\odot}}$
\citep{hig03}.
A disk with the inclination angle of $5\degr$ is impossible, since
the dynamical mass of the central gas is almost the same with
the mass of the whole galaxies.
A disk with the inclination angle of $10\degr$ is still too large,
since more than one tenth of the total mass is concentrated within
a thousandth of radius.
A disk with the inclination angle of $30\degr$ can be possible.
We therefore think that this face-on configuration can be possible,
only if the inclination angle is $\sim30\degr$ or larger.

The latter molecular gas configuration has an edge-on structure,
so the structure should have thin disk- or ring-like structure,
or it can be torus-like structure but has to have a clumpy internal
structure to expose the BLR at the center.
Such clumpy structure is suggested theoretically \citep{wad02}, and
with their model, the column density of $\sim10^{21}$ cm$^{-2}$ is
possible even with the inclination angle of $\sim60\degr$.
Under this model, the difference of the column density derived from
our CO data and from the X-ray data can be explained;
the spatial resolution of our observation is about 250~pc, which
smears all the internal clumpy structures, and therefore the derived
column density will be the average value and higher than that
derived from X-ray observations, which only trace very narrow column
densities
due to the very small size scale of the X-ray emitting region
($<1$~pc).
Under these nearly edge-on configurations, the rotational velocity is
in the typical values for other galaxies ($<500$ km s$^{-1}$), and
therefore we do not need to invoke any special conditions.

Of course, there is another possibility that the central component is
nothing related to the hypothetical torus, namely the central
component we observed is just a part of the galactic disk gas.
This is supported by the similar trend and the smooth connection of
the molecular gas kinematics at the nucleus and the ring.
Since the infalling motion exists from the ring to the nucleus along
the nuclear bar or spiral \citep{pri05,fat06}, it is natural to pile
up the gas around the nucleus with similar kinematics as that of the
ring.
In this case, gas will rotate around the nucleus with similar
inclination as the outer disk or the ring, and it is natural not to
cover the line-of-sight toward the AGN, namely the gas configuration
will be similar to the Seyfert 1 nucleus.
In addition, if the molecular gas piles up at the diameter smaller
than our beam size of 200--300~pc, the discrepancy of the column
density estimated from our data and the X-ray data can also be
explained.
Here we briefly estimate the necessary gas infall rate to create the
central component with gas infall from the molecular ring.
The H$_{2}$ masses of the central component is
$6.5\times10^{7}~{\rm M_{\odot}}$, so that the mass infall rate
of 0.5 M$_\odot$ yr$^{-1}$ is required to create this component
within the molecular ring gas consumption timescale of
$1.2\times10^{8}$ years (Sect.~\ref{sect-dis-ring}).
Note that the line ratio of the central component is obviously
different from that in other regions or other galaxies, and this can
be explained either by the shock caused by the gas inflow along the
nuclear bar or nuclear spiral, or by AGN activities, such as
irradiation of strong X-ray emission to the central component.

In summary, the central component can be the hypothetical
circumnuclear disk or torus with nearly edge-on ($\gtrsim60\degr$)
clumpy structure, or less likely nearly face-on ($\gtrsim30\degr$)
disk/torus.
The central component can also be nothing related to the hypothetical
disk/torus, and possibly created by the gas inflow from the molecular
ring.
The physical conditions of the central component are different from
that of molecular gas in other regions, so that even it is not the
hypothetical disk/torus, it should be highly related to the AGN
activities or gas inflow toward the nucleus.

\section{SUMMARY}

We successfully imaged the central component and the molecular ring,
which is spatially coincident with the AGN and the starburst ring,
toward the central region of the Seyfert 1 galaxy NGC 1097 using the SMA
in the $^{12}$CO(J = 2 -- 1) line.
Here are the summary for the nature of the central component we
observed:
\begin{enumerate}
\item We found that the $^{12}$CO(J = 2 -- 1) map shows an intensity
	peak at the central component, and different from the
	$^{12}$CO(J = 1 -- 0) map, which shows the intensity peak at the
	molecular ring.
\item The $^{12}$CO(J = 2 -- 1)/$^{12}$CO(J = 1 -- 0) line intensity
	ratio for the central component is $1.9\pm0.2$, which is
	different from the global values in GMCs or in spiral galaxies of
	about unity or less.
	From the LVG analysis, we estimated that the central component is
	warmer ($T_{\rm K}\ga400$~K) and denser
	($n_{\rm H_2}\sim3\times10^{4}$~cm$^{-3}$) than that of the
	normal molecular clouds assuming a normal
	$Z(^{12}$CO)/(d$v$/d$r$).
	These values depend highly on the $Z(^{12}$CO)/(d$v$/d$r$), and
	lower $Z(^{12}$CO)/(d$v$/d$r$) for about an order of magnitude
	decreases the temperature and density of about an order of
	magnitude.
	The effect of intense star formation and/or AGN activities, such
	as shocks induced by numerous supernova explosions or strong
	X-ray radiation from AGN are presumably the causes of the unusual
	line ratio.
\item Faster rotation feature of the central gas is observed in
	NGC 1097, which is similar results as observed in other Seyfert 2
	galaxies.
	We interpret this feature as a highly-inclined clumpy disk/torus
	or thin disk to explain the fast rotation and the difference
	between the column density derived from CO and X-ray
	observations.
	A low-inclined ($i\sim30\degr$) thick disk is possible, but lower
	inclination than this value is less likely since it is rare to
	see a velocity $\ge500~{\rm km s}^{-1}$ in nearby galaxies, and
	the total mass of central disk turns to be too large compared
	with the total mass of the galaxy.
	On the other hand, the central component can also be interpret as
	just an inner extention of the galactic disk, possibly created by
	the gas inflow from the molecular ring.
\item Combining kinematical and spatial information with the physical
	conditions, we suggest that the central gas is related to the
	Seyfert activities or gas inflow.
\end{enumerate}
Here are the summary for the nature of the molecular ring we observed:
\begin{enumerate}
\item The $R_{21}$ of the molecular ring is $1.3\pm0.2$, which shows
	a similar properties to the star forming GMCs in our Galaxy or
	nearby starburst galaxies.
	In addition, since the molecular ring shows a good spatial
	coincident with the starburst ring, so we expect that the
	molecular ring is actually fueling the starburst activities.
	The molecular gas mass content with respective to the dynamical
	mass, on the other hand, shows somewhat low value, so that high
	$R_{21}$ may not be related to star formation activities, but
	to other activities, such as shock induced by gas infall along
	the bar.
	Without further replenishment, it can last for only about
	1.2$\times~10^{8}$ years, but since there seems to have gas flows
	from the dust lane along the large-scale bar toward the molecular
	ring, and from the molecular ring toward the nucleus, this
	timescale is highly uncertain.
\end{enumerate}

\acknowledgements

We would like to acknowledge the anonymous referee for the
valuable discussions and comments.
We also thank for the SMA staff for maintaining the operation of the array.
The Submillimeter Array is a joint project between the Smithsonian
Astrophysical Observatory and the Academia Sinica Institute of
Astronomy and Astrophysics and is funded by the Smithsonian
Institution and the Academia Sinica. This work is supported by
the National Science Council (NSC) of Taiwan,
NSC 96-2112-M-001-0095.

\clearpage

\clearpage


\begin{thebibliography}{99}

\bibitem[Antonucci(1993)]{ant93} Antonucci, R.  1993, \araa, 31, 473
\bibitem[Barth et al.(1995)]{barth} Barth, A. J., Ho, L. C.,
        Filippenko, A. V., \& Sargent, W. L.  1995, \aj, 110, 1009
\bibitem[Braine et al.(1993)]{bra93} Braine, J., Combes, F.,
        Casoli, F., Dupraz, C., Gerin, M., Klein, U., Wielebinski, R.,
        \& Brouillet, N.  1993, \aaps, 97, 887
\bibitem[Baker \& Scoville(1998)]{baker98} Baker, A. J. \& Scoville, N. Z. 1998
	, IAUS, 184, 221
\bibitem[Chandler et al.(1996)]{cha96} Chandler, C. J., Terebey, S.,
        Barsony, M., Moore, T. J. T., \& Gautier, T. N.  1996, \apj,
        471, 308
\bibitem[de Vaucouleurs et al.(1991)]{dev91} de Vaucouleurs, G.,
        de Vaucouleurs, A., Corwin, H. G., Jr., Buta, R. J.,
        Paturel, G., \& Fouqu\'e, P.  1991, Third Reference
        Catalogue of Bright Galaxies (New York: Springer-Verlag)
\bibitem[Fathi et al.(2006)]{fat06} Fathi, K., Storchi-Bergmann, T.,
        Riffel, R. A., Winge, C., Axon, D. J., Robinson, A.,
        Capetti, A., \& Marconi, A.  2006, \apjl, 641, L25
\bibitem[Flower \& Launay(1985)]{flo85} Flower, D. R.,
        \& Launay, J. M.  1985, \mnras, 241, 271
\bibitem[Gerin et al.(1988)Gerin, Nakai, \& Combes]{ger88}
        Gerin, M., Nakai, N., \& Combes, F.  1988, \aap, 203, 44
\bibitem[Goldreich \& Kwan(1974)]{gold74} Goldreich, P.,
        \& Kwan, J.  1974, \apj, 189, 441
\bibitem[Hafok \& Stutzki(2003)]{haf03} Hafok, H.,
        \& Stutzki, J.  2003, \aap, 398, 959
\bibitem[Ho et al.(2004)Ho, Moran, \& Lo]{ho04} Ho, P. T. P.,
        Moran, J. M., \& Lo, F.  2004, \apj, 616, L1
\bibitem[Higdon \& Wallin(2003)]{hig03}
        Higdon, J. L. \& Wallin, J. F. 2003, \apj, 585, 281
\bibitem[Hummel et al.(1987)Hummel, van der Hulst, \& Keel]{hum87}
        Hummel, E., van der Hulst, J. M., \& Keel, W. C.  1987,
        \aap, 172, 32
\bibitem[Iyomoto et al.(1996)]{iyo96} Iyomoto, N., Makishima, K.,
        Fukazawa, Y., Tashiro, M., Ishisaki, Y., Nakai, N.,
        \& Taniguchi, Y.  1996, \pasj, 48, 231
\bibitem[Jackson et al.(1993)]{jac93} Jackson, M., Paglione, T. A. D.,
        Ishizuki, S., \& Nguyen-Q-Rieu  1993, \apjl, 418, L13
\bibitem[Keel(1983)]{keel} Keel, W. C.  1983, \apj, 269, 466
\bibitem[Kenney et al.(1992)]{ken92} Kenney, J. D. P., Wilson, C. D.,
        Scoville, N. Z., Devereux, N. A., Young, J. S.  1992,
        \apj, 395, 79
\bibitem[Kohno et al.(1996)]{koh96} Kohno, K., Kawabe, R., Tosaki T.,
        \& Okumura S. K.  1996, \apjl, 461, L29
\bibitem[Kohno et al.(2001)]{koh01} Kohno, K.,  Matsushita, S.,
        Vila-Vilar\'o, B., Okumura, S. K., Shibatsuka, T., Okiura, M.,
        Ishizuki, S., \& Kawabe, R.  2001, in ASP Conf. Ser. 249,
        The Central Kiloparsec of Starbursts and AGN: The La Palma
        Connection, ed. J. H. Knapen, J. E. Beckman, I. Shlosman,
        \& T. J. Mahoney (San Francisco: ASP), 672
\bibitem[Kohno(2005)]{koh05} Kohno, K.  2005, AIP Conf. Proc. 783,
        The Evolution of Starbursts: The 331st Wilhelm and Else Heraeus
        Seminar, ed. S. H\"uttemeister, E. Manthey, D. Bomans, \&
        K. Weis, 203
\bibitem[Kohno et al.(2003)]{koh03} Kohno, K., Ishizuki, S.,
        Matsushita, S., Vila-Vilar\'o, B., \& Kawabe, R.  2003,
        \pasj, 55, L1
\bibitem[Kotilainen et al.(2000)]{kot00} Kotilainen, J. K.,
        Reunanen, J., Laine, S., \& Ryder, S. D.  2000, \aap, 353, 834
\bibitem[Krips et al.(2005)]{kri05} Krips, M., Eckart, A., Neri, R., Pott, J. U.
        , Leon, S., Combes, F., Garc$\acute{\i}$a-Burillo S., Hunt, L. K.,
        Baker, A. J. , Tacconi, L. J., Englmaier, P., Schinnerer, E.,
        \& Boone, F. 2005, \apj, 442, 479
\bibitem[Krips et al.(2007)]{kri07} Krips, M., Neri, R.,
        Garc$\acute{\i}$a-Burillo S.,Combes, F., Schinnerer, E.,
        Baker, A. J., Eckart, A., Boone, F., Hunt, L., Leon, S., \&
        Tacconi, L. J. 2007, \aap, 468, L63
\bibitem[Lavezzi et al.(1999)]{lav99} Lavezzi, T. E., Dickey, J. M.,
        Casoli, F., \& Kaz\`es, I.  1999, \aj, 117, 1995
\bibitem[Lepp \& Dalgarno(1996)]{lepp96} Lepp, S. \& Dalgarno, A.,
	1996, \aap, 306, L21
\bibitem[Lindt-Krieg et al.(2008)]{lindt08} Lindt-Krieg, E., Eckart, A., Neri, R.
        , Krips, M., Pott, J.-U., Garc$\acute{\i}$a-Burillo S.,Combes, F. 2008,
        \apj, 479, 377
\bibitem[Matsushita et al.(1998)]{mat98} Matsushita, S., Kohno, K.,
        Vila-Vilar\'o, B., Tosaki, T., \& Kawabe, R.  1998, \apj,
        495, 267
\bibitem[Matsushita et al.(2004)]{mat04} Matsushita, S., Sakamoto, Kazushi,
        Kuo, Cheng-Yu, Hsieh, Pei-Ying, Dinh-V-Trung, Mao, Rui-Qing,
        Iono, Daisuke, Peck, Alison B., Wiedner, Martina C., Liu, Sheng-Yuan,
        Ohashi, Nagayoshi, \& Lim, Jeremy 2004, \apjl, 616, L55
\bibitem[Mckee et al.(1982)] {lvg-h-t} McKee, C. F., Storey, J. W. V.,
        Watson, D. W., \& Green, S.  1982, \apj, 259, 647
\bibitem[Meier et al.(2000)Meier, Turner, \& Hurt]{mei00}
        Meier, D. S., Turner, J. L., \& Hurt, R. L.  2000, \apj,
        531, 200
\bibitem[MeiJerink et al.(2007) Meijerink, Spaans, \& Israel]{mei07}
	Meijerink, R., Spaans, M., \& Israel, F. P. 2007, \aap, 461, 793
\bibitem[Ondrechen et al.(1989)Ondrechen, van der Hulst, \& Hummel]{ondre89}
        Ondrechen, M. P., van der Hulst, J. M., \& Hummel, E.  1989,
        \apj, 342, 39
\bibitem[Petitpas \& Wilson(2003)]{pet03} Petitpas, G. R.,
        \& Wilson, C. D.  2003, \apj, 587, 649
\bibitem[Prieto et al.(2005)Prieto, Maciejewski, \& Reunanen]{pri05}
        Prieto, M. A., Maciejewski, W. \& Reunanen, J.  2005, \apj,
        130, 1472
\bibitem[Quillen et al.(1995)]{qui95} Quillen, A. C., Frogel, J. A.,
        Kuchinski, L. E., \& Terndrup D. M.  1995, \aj, 110, 156
\bibitem[Rickard(1975)]{rick} Rickard, J. J.  1975, \aap, 40, 339
\bibitem[Richardson et al.(1985)]{ric85} Richardson, K. J.,
        White, G. J., Avery, L. W., Lesurf, J. C. G.,
        \& Harten, R. H.  1985, \apj, 290, 637
\bibitem[Rotaciuc et al.(1991)]{rot91} Rotaciuc, V., Krabbe, A.,
	Cameron, M., Drapatz, S., Genzel, R., Sternberg, A., Storey,
	\& J. W. V., 1991, \apj, 370, 23
\bibitem[Rubin et al.(1980)Rubin, Thonnard, \& Ford]{rubin1}
        Rubin, V. C., Thonnard, N., \& Ford, W. K., Jr. 1980, \apj,
        238, 471
\bibitem[Rubin et al.(1982)]{rubin2} Rubin, V. C., Ford, W. K., Jr.,
        Thonnard, N., \& Burstein, D. 1982, \apj, 261, 439
\bibitem[Sakamoto et al.(1994)]{sak94} Sakamoto, S., Hayashi, M.,
        Hasegawa, T., Handa, T., \& Oka, T.  1994, \apj, 425, 641
\bibitem[Sakamoto et al.(1999)]{sak99} Sakamoto, K., Okumura, S. K.,
        Ishizuki, S., \& Scoville, N. Z.  1999, \apj, 525, 691
\bibitem[Schinnerer et al.(2000) Schinnerer, Eckart, \& Tacconi]{sch2000}
	Schinnerer
        , E., Eckart, A. \& Tacconi, L. J. 2000, \apj, 533, 826
\bibitem[Schinnerer et al.(2000)]{sch00} Schinnerer, E., Eckart, A.,
        Tacconi, L. J., Genzel, R., \& Downes, D.  2000, \apj, 533, 850
\bibitem[Scoville et al.(1987)] {sco87} Scoville, N. Z., Yun, M. S.,
        Clemens, D. P., Sanders, D. B., \& Waller, W. H.  1987,
        \apjs, 63, 821
\bibitem[Scoville et al.(1998)]{sco98} Scoville, N. Z., Yun, M. S.,
        Armus, L., \& Ford, H.  1998, \apjl, 493, L63
\bibitem[Seta et al.(1998)]{set98} Seta, M.,Hasegawa, Tetsuo, Dame, T. M.
	,Sakamoto, Seiichi, Oka, Tomoharu, Handa, Toshihiro, Hayashi, Masahiko,
	Morino, Jun-Ichi, Sorai, Kazuo, \& Usuda, Kumiko S. 1998, \apj, 505, 286
\bibitem[Sofue et al.(1999)]{sofue} Sofue, Y., Tutui, Y., Honma, M., Tomita,
         A., Takamiya, T., Koda, J., \& Takeda, Y. 1999, \apj, 523, 136
\bibitem[Solomon et al.(1987)]{sol87} Solomon, P. M., Rivilo, A. R.,
        Barrett, J., \& Yahil, A.  1987, \apj, 319, 730
\bibitem[Storchi-Bergmann et al.(1993)Storchi-Bergmann, Baldwin, \& Wilson]
        {SB93} Storchi-Bergmann, T., Baldwin, J. A.,
        \& Wilson, A. S.  1993, \apj, 410, L11
\bibitem[Storchi-Bergmann et al.(1996)Storchi-Bergmann, Wilson, \& Baldwin]
        {sto96} Storchi-Bergmann, T., Wilson, A. S.,
        \& Baldwin, J. A.  1996, \apj, 460, 252
\bibitem[Tacconi et al.(1994)]{tac94} Tacconi, L. J., Genzel, R.,
	Blietz, M., Cameron, M., Harris, A. I., \& Madden, S.,
	1994, \apj, 426, 77
\bibitem[Terashima et al.(2002)]{tera02} Terashima, Y., Iyomoto, N.,
        Ho, L. C., \& Ptak, A. F.  2002, \apjs, 139, 1
\bibitem[Tully(1988)]{tully} Tully  1988, Nearby Galaxies Catalog
        (Cambridge: Cambridge University Press)\
\bibitem[Turner et al.(1993)Turner, Hurt, \& Hudson]{tur93}
        Turner, J. L., Hurt, R. L., \& Hudson, D. Y.  1993, \apjl,
        413, L19
\bibitem[Usero et al.(2004)]{usero04} Usero, A., Garc$\acute{\i}$a-Burillo,
	S. Fuente, A.	Mart$\acute{\i}$n-Pintado, J., \&
	Rodr$\acute{\i}$guez-Fern$\acute{\rm a}$ndez, N.~J., 2004, \aap,
	419, 897 
\bibitem[van Dishoek et al.(1993)van Dishoeck, Jansen, \& Phillips]{van93}
        van Dishoeck, E. F., Jansen, D. J., \& Phillips, T. G. 1993,
        \aap, 279, 541
\bibitem[Wada \& Norman(2002)]{wad02} Wada, K.,
        \& Norman, C. A.  2002, \apjl, 566, L21
\bibitem[Weiss et al.(2001)]{wei01} Weiss, A., Neininger, N.,
        H\"uttemeister, S., \& Klein, U.  2001, \aap, 365, 571

\end{thebibliography}
\end{document}